\documentclass[aps,prb,reprint,superscriptaddress,amssymb,amsmath]{revtex4-1}
\usepackage{graphicx}
\usepackage{dcolumn}
\usepackage{bm, color}
\usepackage[none]{hyphenat} 
\usepackage{amsmath}

\begin{document}

\title{Characterization of topological band structures away from the Fermi level by anomalous Nernst measurements}
\author{Jonathan Noky}
\affiliation{Max Planck Institute for Chemical Physics of Solids, D-01187 Dresden, Germany}
\author{Johannes Gooth}
\affiliation{Max Planck Institute for Chemical Physics of Solids, D-01187 Dresden, Germany}
\author{Claudia Felser} 
\affiliation{Max Planck Institute for Chemical Physics of Solids, D-01187 Dresden, Germany}
\author{Yan Sun} 
\email{ysun@cpfs.mpg.de}
\affiliation{Max Planck Institute for Chemical Physics of Solids, D-01187 Dresden, Germany}

\date{\today}

\begin{abstract}
Resolving the structure of energy bands in transport experiments is a major challenge in condensed matter physics and material science. Sometimes, however, traditional electrical conductance or resistance measurements only provide very small signals, and thus limit the ability to obtain direct band structure information. In this case, it has been proven beneficial to employ thermoelectric measurements which are sensitive to the first derivative of the density of states with respect to energy, rather than to its value itself. Due to the large interest in topological effects these days, it is important to identify a similar concept for detecting the Berry curvature in a band structure. Nowadays, the common way to access the Berry curvature directly via measurements is the anomalous Hall effect, but the corresponding signal can be too small to be detected when the topological features of the band structure lie too far off the Fermi level. In this work we propose to investigate topological band structure features utilizing the anomalous Nernst effect which is tied to the derivative of the anomalous Hall effect with respect to energy. Thereby, also signatures become resolvable, which are elusive in anomalous Hall measurements. We demonstrate the underlying mechanisms for a minimal effective four-band model and exemplary for the real Heusler compounds Co$_2$Fe$X$ ($X$=Ge,Sn), which host topological nodal lines about 100 meV above the Fermi level. This work demonstrates that anomalous Nernst measurements can be an effective tool for the characterization of topological band structures, both at room temperature and in the quantum transport regime at cryogenic temperatures.

\end{abstract}
\maketitle
%

Electrical conductance or resistance measurements have been widely used as a powerful tool to probe the energetic band structure of materials. This is because these electrical transport coefficients are directly proportional to the density of states. However, detecting fine structures can be a challenge when the electrical signals become very small. Due to such limitations, thermoelectric measurements have recently been established as a complementary tool to resolve energy bands and various related quantum phenomena. At low temperatures, the thermopower provides a measure of the first derivative of the electrical characteristics with respect to the energy~\cite{mott1969} and is thereby sensitive to the change of the density of states, rather than to its value itself. Therefore, although not containing additional information, the thermoelectric transport coefficients can provide large signals when the electrical transport can hardly be resolved~\cite{gluschke2015characterization}. Hence, as long as the density of states is not constant, the thermopower can give deep insights into the underlying energy structure that would be elusive with other methods.

In addition to the interest in their energetic structure, there has been a focus on the Berry curvature (BC) and the related topological properties of electronic bands over the last few years. The BC is intimately linked to the anomalous Hall conductivity (AHC), which is therefore commonly used to probe topological properties in experiments~\cite{nayak2016,liang2018anomalous}. Following the argumentation above, now the question arises, whether it is possible to also observe small BC effects with an analogous thermoelectric quantity, where the anomalous Hall signals become too small for experimental detection?

This work proposes a method to detect small signatures of BC utilizing the anomalous Nernst effect (ANE)~\cite{Nerst_1887,Lee_2004, XiaoDi_2006,WBauer2012,Nakatsuji2015,Kiyohara_2016,PhysRevLett.119.056601,PhysRevLett.118.136601,Ikhlas_2017,liang2018anomalous}, which can be understood as the first derivative of the anomalous Hall effect (AHE) with respect to energy~\cite{mott1969,XiaoDi_2006, Xiao2010}. It is shown that topological features in the band structure, which are not close enough to the Fermi level to have a detectable influence on the AHC, can give rise to large anomalous Nernst signals. Compared to the direct relation between the energy structure and the thermopower, the relation between topological properties and the ANE is more subtle. In one way, one may summarize the connection as follows: The ANE is related to the first derivative of the AHE; and the BC affects the AHE; hence, Berry curvature-effects can show up in Nernst signals. In the following, we start with the discussion of a minimal effective four-band model to get a deeper conceptual understanding of the above-mentioned effect. Subsequently, we exemplary present results for a real material, the Heusler compounds Co$_2$Fe$X$ ($X$=Ge,Sn) which exhibit only very small AHC but a strong anomalous Nernst conductivity (ANC). The ANE in these compounds allows to be sensitive to a topological nodal line structure approximately 100 meV above the Fermi level.

%
\begin{figure*}[htb]
\centering
\includegraphics[width=0.96\textwidth]{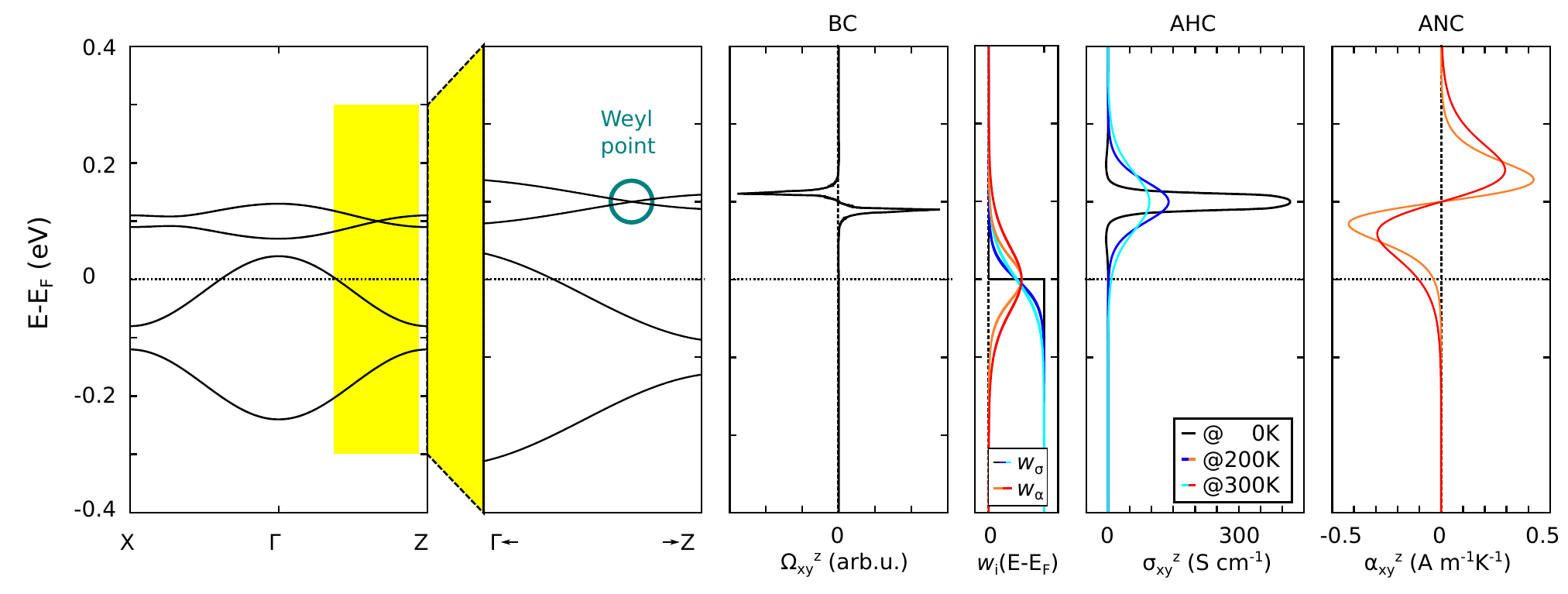}
   \caption{Results for the minimal effective model described in the main text. From left to right: The band structure (left) with Weyl points above the Fermi level (next panel shows the yellow part enlarged). Berry curvature at fixed energy integrated over the whole Brillouin zone. Weighting functions for the anomalous Hall ($w_\sigma$) and Nernst ($w_\alpha$) conductivity. Anomalous Hall conductivity with a zero (very small for finite temperature) value at the Fermi level. Anomalous Nernst conductivity with a finite value at the Fermi level.}
\label{fig:model}
\end{figure*}
For the investigation we employ \textit{ab initio} calculations based on density-functional theory (DFT) as implemented in the \textsc{VASP} package~\cite{kresse1996}. Plane waves with pseudopotentials are used as a basis set and for the description of the exchange-correlation potential we utilize the generalized-gradient approximation (GGA)~\cite{perdew1996}. From this electronic structure we extract Wannier functions using the \textsc{Wannier90} package~\cite{Mostofi2008} with initial projections to the s-, p-, and d- orbitals of Co and Fe and to the s- and p-orbitals of $X$. We construct a Tight-Binding Hamiltonian $H$ from the Wannier functions to evaluate the BC $\Omega$ in the system via the Kubo formula~\cite{Thouless_1982,Xiao2010,Nagaosa_2010}
\begin{equation}
  \Omega_{ij}^n=\sum_{m\ne n} \frac{\langle n|\frac{\partial H}{\partial k_i}|m\rangle \langle m|\frac{\partial H}{\partial k_j}|n\rangle - (i \leftrightarrow j)}{(E_n-E_m)^2},
\end{equation}
with $|n\rangle$ and $E_n$ being the eigenstates and -energies of $H$.

We now calculate the AHC $\sigma$ from the BC via
\begin{equation}
 \label{eq:ahc}
 \sigma_{ij}=\frac{e^2}{\hbar} \sum_n\int \frac{d^3k}{(2\pi)^3}\Omega_{ij}^nf_n
\end{equation}
and the ANC $\alpha$ using the equation proposed by Xiao et al. \cite{Xiao2010, XiaoDi_2006}
\begin{align}
  \label{eq:anc}
  \alpha_{ij}=-\frac{1}{T} \frac{e}{\hbar} \sum_n \int \frac{d^3k}{(2\pi)^3} \Omega_{ij}^{n}[(E_{n}-E_F)f_{n}+\nonumber \\
  +k_BT\ln{(1+\exp{\frac{E_{n}-E_F}{-k_BT}})}].
\end{align}
Here, $T$ is the actual temperature, $f_n$ is the Fermi distribution, and $E_F$ is the Fermi level.
%

As it can be seen from the above equations, both AHC and ANC can be written in an analogous form as 
\begin{equation}
\label{eq:general}
  \lambda_{ij}=\frac{e^2}{\hbar}\sum_n\int \frac{d^3k}{(2\pi)^3}\Omega_{ij}^nw_\lambda(E_n-E_F) \text{ with } \lambda=\sigma,\alpha.
\end{equation}
So both effects stem from the BC of the bands but with different weighting factors $w$, which are
\begin{align}
\label{eq:hallfac}
  w_\sigma(E)&=f_n^{T}(E)\\
  \label{eq:nernstfac}
  w_\alpha(E)&=-\frac{1}{eT}[Ef_{n}^T+k_BT\ln{(1+\exp{\frac{E}{-k_BT}})}],
\end{align}
where $f_n^T$ is the Fermi distribution for the temperature $T$. The distribution functions are shown in the fourth graph in Fig.~\ref{fig:model}. Here, a significant difference between AHE and ANE becomes obvious. On the one hand, to achieve a large AHC and ANC, strong BC is needed as it can be seen from equation (\ref{eq:general}). However, on the other hand the two effects differ in the detailed weights of the contributions $w$ (see equations (\ref{eq:hallfac}) and (\ref{eq:nernstfac})). The AHC arises from the BC of all occupied bands while the ANC stems from both occupied and unoccupied bands but only near the Fermi level. Here, also the influence of temperature $T$ becomes obvious, as with higher $T$ the weighting $w_\alpha$ gets broader, including states further away from $E_F$. This leads to quite distinct behaviours.

In the following we discuss a minimal effective four-band model derived from the one proposed in Refs.~\cite{Lu_2015,Noky2018}.
The Hamiltonian can be written as
\begin{align}
 H&=\begin{bmatrix}
     \hspace{-0.8cm}M_0\sigma_0+tB(\vec{k})\sigma_z+\alpha A(\vec{k})\qquad\qquad 0\qquad\qquad \\
     \qquad0 \qquad-M_0\sigma_0+(t'B(\vec{k})+\delta)\sigma_z+\alpha' A(\vec{k})
    \end{bmatrix}\nonumber\\
 &\text{with }A(\vec{k})=\sin{k_x}\sigma_x+\sin{k_y}\sigma_y \nonumber \\
 &\text{and }B(\vec{k})=\cos{k_x}+\cos{k_y}+\cos{k_z}, \nonumber
\end{align}
where $\sigma$ are the Pauli matrices for spin and $M_0$, $\alpha$, $\alpha'$, $t$, $t'$, and $\delta$ are model parameters. As model parameters we used $M_0=0.1\:\text{eV}$, $\alpha=0\:\text{eV}$, $\alpha'=0.005\:\text{eV}$, $t=0.03\:\text{eV}$, $t=0.01\:\text{eV}$, and $\delta=0.05\:\text{eV}$. This leads to a system with Weyl points around 100 meV above $E_F$ and with only a trivial band at $E_F$ (see Fig.~\ref{fig:model} left panel). For this band structure the BC is shown in the third panel of Fig.~\ref{fig:model}. Here, the Weyl points are the only source of BC in the system and the BC induced by them decays to zero before reaching $E_F$. Multiplying the BC distribution with the weighting functions $w_\sigma$ and $w_\alpha$ results in the AHC and ANC, which are shown in the right panels of Fig.~\ref{fig:model}. It can be seen that the AHC peaks at the energy where Weyl points are located and reaches zero at the $E_F$ for $T=0$ K. With increasing temperature the AHC still is very weak at $E_F$. However, due to the distribution of the BC in combination with $w_\alpha$ the ANC still has a nonzero value at $E_F$, where the value itself is also temperature dependent due to the temperature dependence of $w_\alpha$. 

Remarkably, this behaviour can not be understood in terms of the Mott relation~\cite{XiaoDi_2006, Xiao2010}
\begin{equation}
  \label{eq:mott}
  \alpha_{ij} =  \frac{\pi^2}{3} \frac{k_B^2 T}{e} \frac{\partial\sigma_{ij}}{\partial E}(E_F),
\end{equation}
which links the ANC to the derivative of the AHC at $T=0$ K with respect to energy. At $E_F$ the AHC is constantly zero but nonetheless a non-zero ANC appears. This model captures the different mechanisms of the AHE and ANE and the possibility to see topological features located away from the Fermi level via the ANE.

\begin{figure}[htb]
\centering
\includegraphics[width=0.48\textwidth]{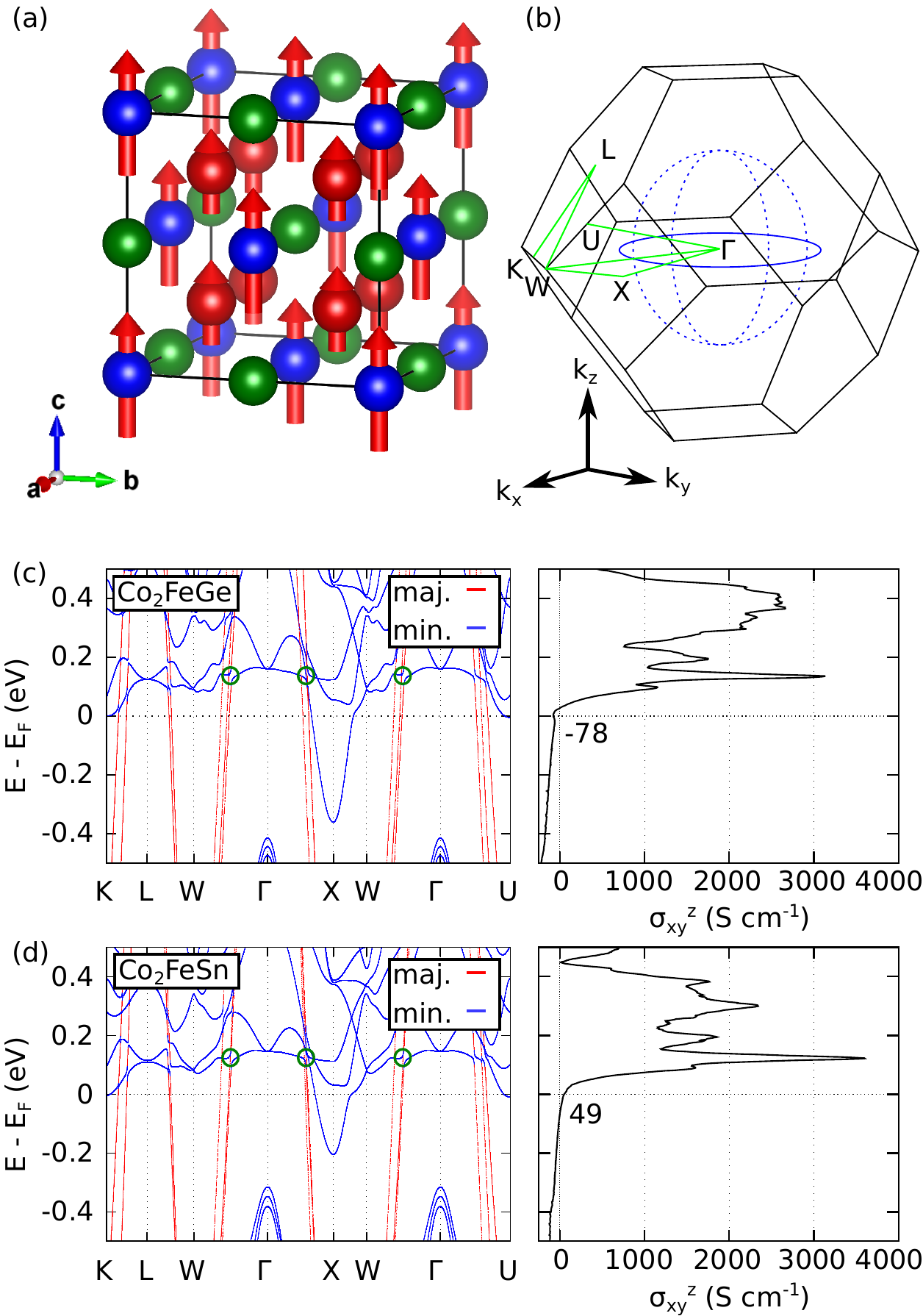}
   \caption{(a) Crystal structure of Co$_2$Fe$X$ ($X$=Ge,Sn) with magnetic moments aligned along the (001) direction. (b) Brillouin zone with the high-symmetry points for the band structure (green) and the schematically drawn nodal lines (blue). The dashed lines become gapped due to the spin-orbit coupling and magnetization direction. (c) Band structure with spin contributions and anomalous Hall conductivity of Co$_2$FeGe. (d) Band structure with spin contributions and anomalous Hall conductivity of Co$_2$FeSn. The green circles mark the positions of the nodal lines.}
\label{fig:material}
\end{figure}

The effects described in the above model can be seen in the real meterials Co$_2$Fe$X$ ($X$=Ge,Sn). These compounds are regular full Heusler compounds with space group $Fm\overline{3}m$ (No. 225). The crystal structure is depicted in Fig.~\ref{fig:material}(a), where also the magnetic moments of the atoms are shown. The compounds show a ferromagnetic behaviour with a net magnetic moment of $\mu=5.6\mu_B$ per unit cell aligned along the (001) direction. The moments of the single atoms are $\mu_{Co}=1.4\mu_B$, $\mu_{Fe}=2.9\mu_B$, and $\mu_X=0$. Both compounds host an interesting nodal line structure which is shown in the Brillouin zone in Fig.~\ref{fig:material}(b). There are three lines approximately 100 meV above the Fermi level (see Fig.~\ref{fig:material}(c) and (d)) in the main coordinate planes. Due to the magnetization direction and the spin-orbit coupling two of them become gapped (dashed lines in Fig.~\ref{fig:material}(b)). The third line is still closed because of the mirror symmetry which still holds in this plane. Changing the magnetization direction will also change the gapping behaviour of the nodal lines without influencing the transport properties. This behaviour was also recently observed in Co$_2$MnGa~\cite{Guin2018}. The band structure and the AHC for both compounds look similar (see Fig.~\ref{fig:material}(c) and (d)) and exhibit a large peak structure in the AHC near the nodal lines, where also many other flat d-bands are located. At the Fermi level the AHC is very small (-78 S cm$^{-1}$ and 49 S cm$^{-1}$ for Co$_2$FeGe and Co$_2$FeSn, respectively). These results for the AHC are in good agreement with previous theoretical studies~\cite{Huang2015}. Also, taking the influence of temperature into account the AHC in both compounds changes by less then 100 S cm$^{-1}$ when raising $T$ up to 300 K. Therefore, the topological features of the band structure in these systems, which are located approximately 100 meV above $E_F$, can hardly be detected via the AHE.

\begin{figure}[htb]
\centering
\includegraphics[width=0.48\textwidth]{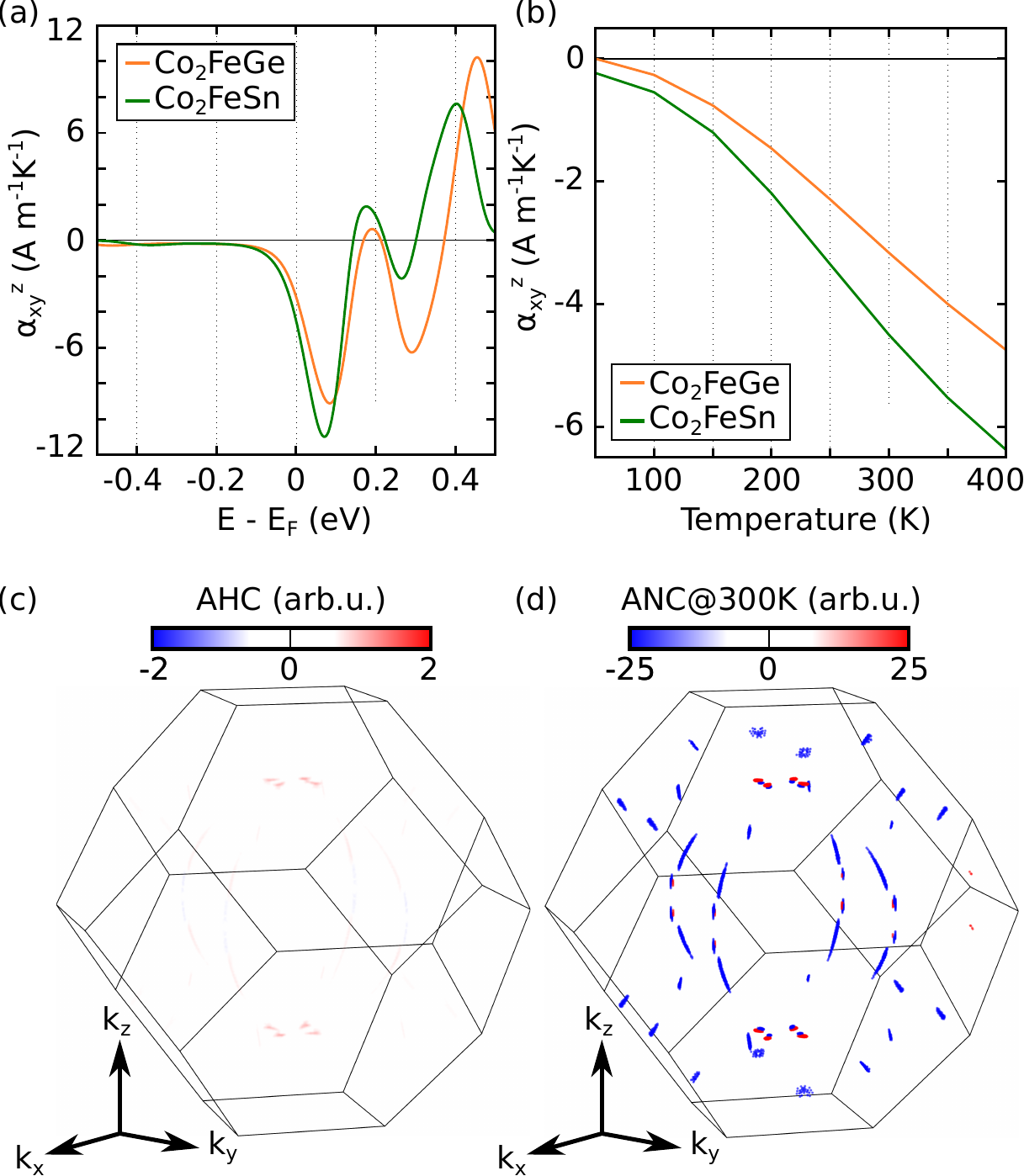}
   \caption{(a) Anomalous Nernst effect of Co$_2$FeGe and Co$_2$FeSn in dependence of the Fermi level. (b) Temperature dependence of the anomalous Nernst effect. (c)+(d) Distribution of the anomalous Hall (c) and Nernst ($T= 300$ K) (d) conductivity of Co$_2$FeSn in the Brillouin zone. Co$_2$FeGe shows a similar behaviour.}
\label{fig:anct}
\end{figure}

\begin{figure}[htb]
\centering
\includegraphics[width=0.48\textwidth]{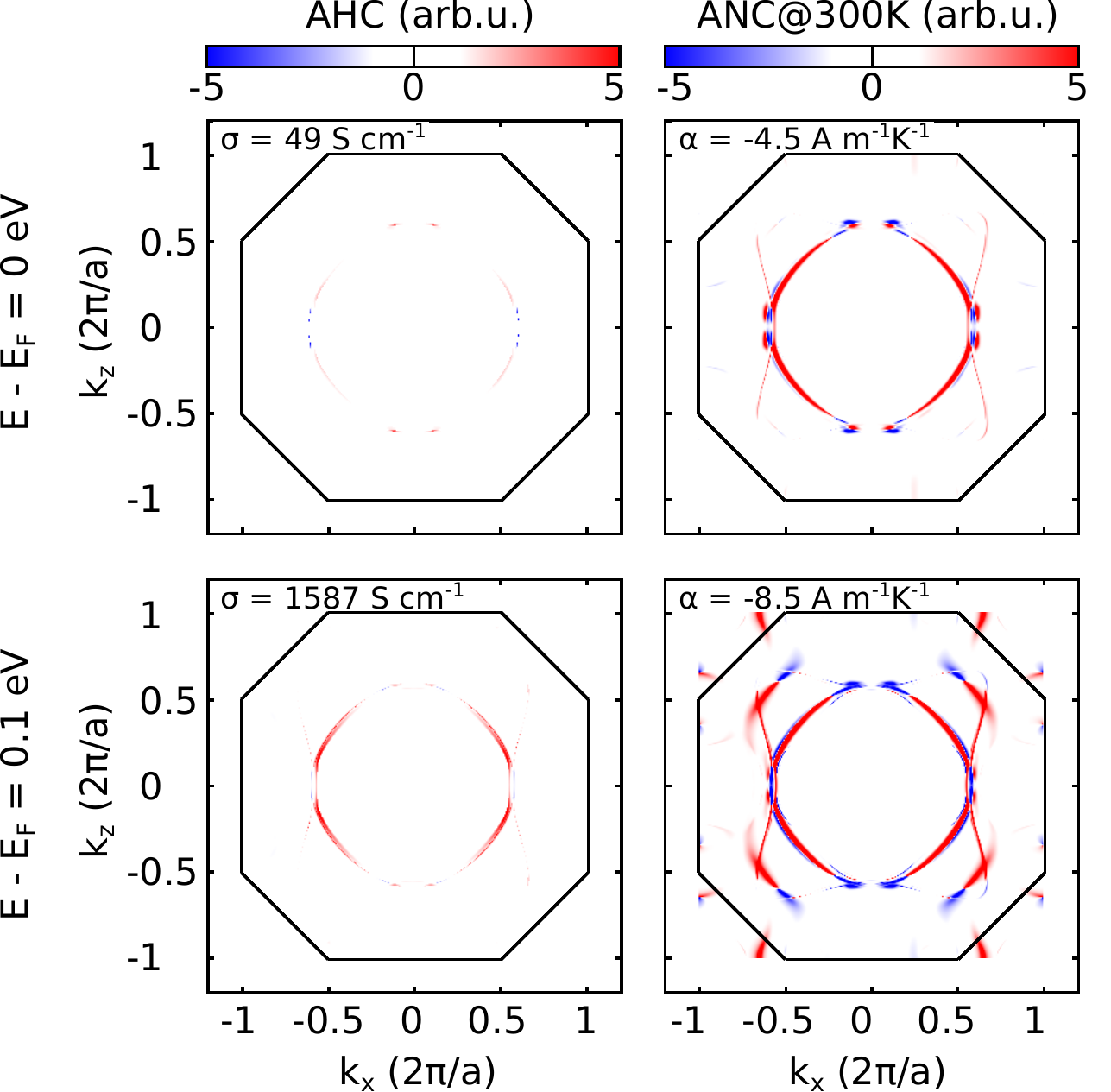}
   \caption{AHC and ANC at 300 K in a cut through the Brillouin zone at $k_y=0$ for different energies levels. One of the gapped nodal lines lies in this plane. The color scale for AHC and ANC is the same for the different energy levels.}
\label{fig:fig4}
\end{figure}
As discussed above for the model system, which also hosts Weyl points 100 meV above $E_F$, it should be possible to probe the topological band structure of Co$_2$Fe$X$ ($X$=Ge,Sn) via the ANE. Indeed, looking at the ANC of the compounds in Fig.~\ref{fig:anct}(a), it turns out to be large at $E_F$. Due to the different weighting functions $w_\sigma$ and $w_\alpha$ the BC induced by the close-lying bands and the nodal lines above $E_F$ influences the ANC much more than the AHC. As the distribution of $w_\alpha$ shows, in contrast to $w_\sigma$ also unoccupied states are contributing to it. Because the topologically interesting effects are located above $E_F$, they only contribute to the ANE. This interpretation is also supported by looking at the temperature dependence in Fig.~\ref{fig:anct}(b). Here, a monotonous increase of the ANC with temperature is visible which can be understood in terms of the broadening of $w_\alpha$ with increasing $T$. This broadening enables more of the BC induced above the Fermi level to contribute. 

As AHC and ANC show a quite different behaviour in these compounds we analysed their distribution in the \textit{k} space. In Fig.~\ref{fig:anct}(c) and (d) the contributions to the AHC and ANC of Co$_2$FeSn, respectively, are displayed (the results for Co$_2$FeGe show a similar behaviour). It is visible that the gapped nodal lines (see Fig.~\ref{fig:material}(b)) are a source of BC in the system, similarly to the behaviour recently observed in Co$_2$MnGa~\cite{Guin2018}. As it can be seen the value in the $k_x$-$k_y$ plane is zero. The reason for this is, that there the nodal line is still gapless because the plane is also a mirror plane, which is linked to the magnetization direction along (001). As it can be seen from the AHC in Fig.~\ref{fig:material}(c) and (d) the AHC is very weak in the whole Brillouin zone. However, looking at the contributions to the ANC in Fig.~\ref{fig:anct}(d), most of the points on the gapped nodal lines now contribute strongly. Additionally, there are spots away from these lines with a high ANC contribution. These points are other band touchings from the flat bands also about 100 meV above the Fermi level. From this results it can already be seen, that the ANE is capable of detecting topological band structure features that are not visible in the AHE. 

To get a deeper insight into this, we now investigate the plane in $k$ space with $k_y=0$. In this plane one of the gapped nodal lines is located. In Fig.~\ref{fig:fig4} we compare the AHC and the ANC for two different energies: at $E_F$ and 100 meV above. For the AHC it can be seen that at the Fermi level the effect is very weak but increases by almost two orders of magnitude when raising the energy up to the position of the nodal lines. However, the ANC already takes a large value of 4.5 A m$^{-1}$K$^{-1}$ at $E_F$ and almost doubles when reaching the higher energy. By comparing the intensities of AHC and ANC in this special $k$ plane it can be concluded that the BC at a higher energy already influences the ANE at lower energies, making it an interesting tool to probe band structures for topological features.
%

In summary, we have investigated a method to probe the topological features of band structures utilizing anomalous thermoelectric transport. This method is especially interesting because it can detect features that are too far away from the Fermi level to influence the anomalous Hall effect. To demonstrate this, we have theoretically investigated the ferromagnetic regular Heusler compounds Co$_2$Fe$X$ ($X$=Ge,Sn). These compounds host a nodal line structure with strong Berry curvature around 100 meV above the Fermi level. Due to the distance to $E_F$ it can not be detected in the anomalous Hall conductivity, which consequently is almost zero. However, it leads to a strong anomalous Nernst conductivity at the Fermi level, which we explained via a detailed analysis of the underlying mechanisms of the two effects in a real system and a minimal effective model. This is also reflected in the temperature dependence of the ANC. Our work shows for the first time the possibility to probe hidden topological features utilizing thermoelectric effects and paves the way for investigating band structures which host topological features away from the Fermi level.

This work was financially supported by the ERC Advanced Grant No. 291472 'Idea Heusler' and ERC Advanced Grant No. 742068 'TOPMAT'.

\end{document}